\documentclass[11pt]{llncs}
\usepackage{amsmath,amssymb,amsbsy,amsfonts,latexsym,
               amsopn,amstext,amsxtra,euscript,amscd,enumitem}

\usepackage{color}

\newcommand{\tr}[2][1]{Tr_{#1}^{#2}}

\newcommand{\GF}[2][2]{{\mathbb F}_{#1^{#2}}}

\def\urltilda{\kern -.15em\lower .7ex\hbox{\~{}}\kern .04em}

\newtheorem{thm}{Theorem}[section]
\numberwithin{thm}{section}

\newtheorem{cor}[thm]{Corollary}

\newtheorem{rem}[thm]{Remark}

\def\urltilda{\kern -.15em\lower .7ex\hbox{\~{}}\kern .04em}

\begin{document}
\title{Characterizations of o-polynomials by the Walsh transform}
\author{Claude Carlet and Sihem Mesnager }\institute{LAGA, Department of Mathematics, University of Paris 8 \\(and Paris 13 and CNRS), Saint--Denis cedex 02, France.\\E-mail: \email{claude.carlet@univ-paris8.fr, smesnager@univ-paris8.fr}}

\date{\today}
\maketitle

\begin{abstract}
  The notion of o-polynomial comes from finite projective
  geometry. In 2011 and later, it has been shown that those objects play an
  important role in symmetric cryptography and coding theory to design
  bent Boolean functions, bent vectorial Boolean functions, semi-bent functions and to construct good linear
  codes. In this note, we characterize o-polynomials by the Walsh
  transform of the associated vectorial functions.
\end{abstract}

{\bf Keywords:} o-polynomial, vectorial function, Walsh--Hadamard transform.

\section{Introduction}In all this paper, $n$ is a positive integer.
The projective space $PG(2,{\Bbb F}_{2^n})$ is a point-line incidence
structure whose points are the one-dimensional vector subspaces of
${\Bbb F}_{2^n}^3$, whose lines are the two-dimensional vector
subspaces of ${\Bbb F}_{2^n}^3$, and where a point is incident with a line
if and only if the one-dimensional vector subspace corresponding to
the point is a subspace of the two-dimensional vector subspace
corresponding to the line. A set of $k$ points such that no three of them lie
in a common line is called a $k$-arc.  The maximum cardinality $k$ of an
arc in $PG(2,2^n)$ is $2^n+2$.  An \emph{oval} of $PG(2,2^n)$ is an
arc of cardinality $2^n+1$. A \emph{hyperoval} of $PG(2,2^n)$ is an
arc of maximum cardinality $2^n+2$. A hyperoval $\mathcal O$ of
$PG(2,2^n)$ containing the fundamental quadrangle (that is, the set of points $(1,0,0)$, $(0,1,0)$, $(0,0,1)$ and $(1,1,1)$) can be described as
$\mathcal O=\{(1,t,G(t)), t\in \mathbb{F}_{2^m}\}\cup\{(0,1,0),
(0,0,1)\}$ for some polynomial $G$ that is called an
\emph{o-polynomial}. Those polynomials can be characterized as follows:
\begin{thm}\label{o-polynome}(\cite{Hirschfeld98})
  A polynomial $G$ over ${\Bbb F}_{2^n}$ is an $o$-polynomial if and only
  $G$ is a permutation and each of the polynomials $G_s$, $s\in{\Bbb F}_{2^n}$,
  is a permutation where
  \begin{displaymath}
    G_s(t) = \left\{\begin{array}{ll}
                    \displaystyle\frac{G(t+s)+G(s)}{t} & \mbox{if $t\not=0$}\\
                      0 & \mbox{if $t=0$.}
                  \end{array}\right.
  \end{displaymath}
\end{thm}

Recently, in \cite{CarletMesnagerClassH2011}, has been discovered a
relation between Niho bent functions and o-polynomials when $n$ is
even. Niho bent functions are those Boolean functions over ${\Bbb F}_{2^n}$ whose
restrictions to ${\Bbb F}_{2^{\frac n2}}$ are linear.  It has
been shown that each Niho bent function corresponds to an
$o$-polynomial and that each $o$-polynomial gives rise to Niho bent
functions. Further, it has been shown in \cite{MesnagerIMA2013}, that
o-polynomials also give rise to infinite families of semi-bent
functions in even dimension. Very recently, new connections between bent vectorial functions and the hyperovals of the
projective plane (extending the link provided in \cite{CarletMesnagerClassH2011} between bent Boolean functions and the hyperovals) as well as a connection between 
o-polynomials in even characteristic and $2^r$-ary simplex codes have
been established in \cite{MesnagerDCC2015}. Furthermore, Ding has provided a nice article \cite{Ding2016} 
where o-polynomials are employed to construct linear codes.

In this note, we present a new characterization of $o$-polynomials by
means of their Walsh transform.

\section{Preliminaries}\label{prel}
Given a finite set $E$, $\vert E\vert$ denotes the cardinality of $E$.
Recall that  for any positive integers $k$, and $r$
dividing $k$, the trace function from $\GF{k}$ to $\GF{r}$, denoted
by $\tr[r]{k}$, is the mapping defined for every $x\in\GF k$ as:
\begin{displaymath}
  \tr[r]{k}(x):=\sum_{i=0}^{\frac kr-1}
  x^{2^{ir}}=x+x^{2^r}+x^{2^{2r}}+\cdots+x^{2^{k-r}}.
\end{displaymath}
  
In particular, the {\em absolute trace} occurs for $r=1$. We shall
denote $tr_n$ the absolute trace function $ \tr[1]{n}$.  Let $m$ be a positive integer and
$F: {\Bbb F}_{2^n}\rightarrow {\Bbb F}_{2^m}$ be a vectorial Boolean
function. The Walsh transform of $F$ at
$(u,v)\in {\Bbb F}_{2^n}\times{\Bbb F}_{2^m}$ equals  by definition the Walsh transform
of the so-called component function $tr_m(v(F(x))$ at $u$, that is:
$$W_F(u,v):=\sum_{x\in {\Bbb F}_{2^n}}(-1)^{tr_m(v(F(x))+tr_n(ux)}.$$

\section{A characterization of o-polynomials}
Consider a polynomial over ${\Bbb F}_{2^n}$ and its associated
function from ${\Bbb F}_{2^n}$ to ${\Bbb F}_{2^n}$. We shall denote
the polynomial by $F(X)$ and the function by $F(x)$. It is well-known (see e.g. \cite{CarletMesnagerClassH2011})
that $F(X)$ is an $o$-polynomial if and only if, for every $a$ and
$b\neq 0$ in ${\Bbb F}_{2^n}$, the equation $F(x)+bx=a$ has 0 or 2
solutions (and we know that $F$ is then a permutation).  Consider the
polynomial over ${\Bbb R}$ equal to $X(X-2)^2=X^3-4X^2+4X$. It takes
value 0 when $X$ equals 0 or 2 and takes strictly positive value when
$X$ is in ${\Bbb N}\setminus \{0,2\}$. We have then that
$$\left|\{x\in {\Bbb F}_{2^n}; F(x)+bx+a=0\}\right|^3-4\left|\{x\in
  {\Bbb F}_{2^n}; F(x)+bx+a=0\}\right|^2$$$$+4\left|\{x\in {\Bbb
    F}_{2^n}; F(x)+bx+a=0\}\right|\geq
0,$$and
$F(X)$ is an o-polynomial if and only if this inequality is an
equality for every nonzero $b\in {\Bbb F}_{2^n}$ and every $a\in {\Bbb
  F}_{2^n}$. Equivalently, we have $$\sum_{a,b\in {\Bbb F}_{2^n},
  b\neq 0}{\Big (}\left|\{x\in {\Bbb F}_{2^n};
  F(x)+bx+a=0\}\right|^3-4\left|\{x\in {\Bbb F}_{2^n};
  F(x)+bx+a=0\}\right|^2$$$$+4\left|\{x\in {\Bbb F}_{2^n};
  F(x)+bx+a=0\}\right|{\Big )}\geq 0,$$ and $F(X)$ is an o-polynomial
if and only if this inequality is an equality.

We shall now characterize this condition by means of the Walsh transform.
We have: $$\left|\{x\in {\Bbb F}_{2^n}; F(x)+bx+a=0\}\right|=2^{-n}\sum_{x\in {\Bbb F}_{2^n}, v\in {\Bbb F}_{2^n}}(-1)^{tr_n(v(F(x)+bx+a))},$$and therefore, for $j\geq 1$:$$\sum_{a,b\in {\Bbb F}_{2^n}, b\neq 0}\left|\{x\in {\Bbb F}_{2^n}; F(x)+bx+a=0\}\right|^j=$$$$2^{-jn}\sum_{a,b\in {\Bbb F}_{2^n}, b\neq 0}\sum_{x_1,\dots ,x_j\in {\Bbb F}_{2^n}\atop v_1,\dots ,v_j\in {\Bbb F}_{2^n}}(-1)^{\sum_{i=1}^jtr_n(v_i(F(x_i)+bx_i+a))},$$and using that $\sum_{a\in {\Bbb F}_{2^n}}(-1)^{tr_n(va)}$ is nonzero for $v=0$ only and takes then value $2^n$, we deduce that $$\sum_{a,b\in {\Bbb F}_{2^n}, b\neq 0}\left|\{x\in {\Bbb F}_{2^n}; F(x)+bx+a=0\}\right|^j=$$$$2^{-(j-1)n}\sum_{b\in {\Bbb F}_{2^n}^*}\sum_{v_1,\dots ,v_j\in {\Bbb F}_{2^n}\atop \sum_{i=1}^j v_i=0}\prod_{i=1}^jW_F(bv_i,v_i)$$ and that $$\sum_{a,b\in {\Bbb F}_{2^n}, b\neq 0}{\Big (}\left|\{x\in {\Bbb F}_{2^n}; F(x)+bx+a=0\}\right|^3-4\left|\{x\in {\Bbb F}_{2^n}; F(x)+bx+a=0\}\right|^2$$$$+4\left|\{x\in {\Bbb F}_{2^n}; F(x)+bx+a=0\}\right|{\Big )}=$$$$2^{-2n}\sum_{b\in {\Bbb F}_{2^n}^*}\sum_{v_1,v_2\in {\Bbb F}_{2^n}}W_F(bv_1,v_1)W_F(bv_2,v_2)W_F(b(v_1+v_2),v_1+v_2)$$$$-2^{-n+2}\sum_{b\in {\Bbb F}_{2^n}^*}\sum_{v\in {\Bbb F}_{2^n}}W_F^2(bv,v)+(2^n-1)2^{n+2}.$$
We have $\sum_{b\in {\Bbb F}_{2^n}^*}\sum_{v\in {\Bbb F}_{2^n}}W_F^2(bv,v)=(2^n-1)W_F^2(0,0)+\sum_{u\in {\Bbb F}_{2^n}^*, v\in {\Bbb F}_{2^n}^*}W_F^2(u,v)=(2^n-1)2^{2n+1}-\sum_{v\in {\Bbb F}_{2^n}^*}W_F^2(0,v)$.
We deduce:
\begin{thm}\label{thmfond}
Let $F$ be any $(n,n)$-function and $F(X)$ be the associated polynomial over ${\Bbb F}_{2^n}$.  Then:
$$\sum_{b\in {\Bbb F}_{2^n}^*}\sum_{v_1,v_2\in {\Bbb F}_{2^n}}W_F(bv_1,v_1)W_F(bv_2,v_2)W_F(b(v_1+v_2),v_1+v_2)$$\begin{equation}\label{ineqthmfond}+2^{n+2}\sum_{v\in {\Bbb F}_{2^n}^*}W_F^2(0,v)- 2^{4n+2}+2^{3n+2}\geq 0,\end{equation}
and this inequality is an equality if and only if $F$ is an o-polynomial.
\end{thm}
The sum $\sum_{b\in {\Bbb F}_{2^n}}\sum_{v_1,v_2\in {\Bbb F}_{2^n}}W_F(bv_1,v_1)W_F(bv_2,v_2)W_F(b(v_1+v_2),v_1+v_2)$ is equal to $$2^n\sum_{v_1,v_2\in {\Bbb F}_{2^n}}\sum_{(x_1,x_2,x_3)\in {\Bbb F}_{2^n}^3\atop v_1x_1+v_2x_2+(v_1+v_2)x_3=0} (-1)^{v_1F(x_1)+v_2F(x_2)+(v_1+v_2)F(x_3)}.$$ We deduce:
\begin{cor}Let $F$ be any $(n,n)$-function and $F(X)$ be the associated polynomial over ${\Bbb F}_{2^n}$.  Then:
$$2^n\sum_{v_1,v_2\in {\Bbb F}_{2^n}}\sum_{(x_1,x_2,x_3)\in {\Bbb F}_{2^n}^3\atop v_1x_1+v_2x_2+(v_1+v_2)x_3=0} (-1)^{v_1F(x_1)+v_2F(x_2)+(v_1+v_2)F(x_3)}$$$$-\sum_{v_1,v_2\in {\Bbb F}_{2^n}}W_F(0,v_1)W_F(0,v_2)W_F(0,v_1+v_2)$$\begin{equation}\label{ineqthmfond}+2^{n+2}\sum_{v\in {\Bbb F}_{2^n}^*}W_F^2(0,v)- 2^{4n+2}+2^{3n+2}\geq 0,\end{equation}
and this inequality is an equality if and only if $F$ is an o-polynomial.
\end{cor}
\begin{rem}
Observe that, when $F$ is an $o$-polynomial, we have:
\begin{eqnarray*}
  \sum_{v\in{\Bbb F}_{2^n}} W^2_F(bv,v) &=& \sum_{v\in{\Bbb F}_{2^n}}\sum_{x_1,x_2\in{\Bbb F}_{2^n}} (-1)^{tr_n(v(F(x_1)+F(x_2)+b(x_1+x_2)}\\
                                        &=&  \sum_{v\in{\Bbb F}_{2^n}}\sum_{\gamma,z \in{\Bbb F}_{2^n}} (-1)^{tr_n(v(F(z+\gamma)+F(\gamma)+bz)}\\
  &=& 2^n\sum_{\gamma\in{\Bbb F}_{2^n}} \vert\{z\in{\Bbb F}_{2^n}\mid F(z+\gamma)+F(\gamma)+bz=0\}\vert.
\end{eqnarray*}
Now,
$\vert\{z\in{\Bbb F}_{2^n}\mid F(z+\gamma)+F(\gamma)+bz=0\}\vert=2$
for every $\gamma\in{\Bbb F}_{2^n}$ according to Theorem
\ref{o-polynome}. And thus, if $F$ is an $o$-polynomial,
\begin{eqnarray*}
   \sum_{v\in{\Bbb F}_{2^n}} W^2_F(bv,v) = 2^{2n+1}.
\end{eqnarray*}
One can then deduce from Theorem \ref{thmfond} that
A polynomial $F(X)$ is an $o$-polynomial if and only if
\begin{eqnarray*}
  \sum_{b\in{\Bbb F}_{2^n}^\star}\sum_{v\in{\Bbb F}_{2^n}} W^2_F(bv,v) = (2^n-1)2^{2n+1}.
\end{eqnarray*}
and
\begin{eqnarray*}
  \sum_{b\in {\Bbb F}_{2^n}^*}\sum_{v_1,v_2\in {\Bbb F}_{2^n}}W_F(bv_1,v_1)W_F(bv_2,v_2)W_F(b(v_1+v_2),v_1+v_2)=(2^n-1)2^{3n+2}.
\end{eqnarray*}
\end{rem}


\def\cprime{$'$} \def\cprime{$'$}

\end{document}